\documentclass[a4paper,12pt]{article}

\title{Eternally inflating cosmologies from intersecting spacelike branes}
\author{Riuji Mochizuki\thanks{e-mail address:rjmochi@tdc.ac.jp}
 \\  {\small  Laboratory of Physics, Tokyo Dental College, Chiba 261-8502, Japan}}

\date{\today}

\begin{document}

\maketitle
\abstract{

Intersecting spacelike braneworld cosmologies are investigated.  
The time axis is set on the scale parameter of extra space, which may include more than one timelike metric.   
Obtained are eternally inflating (i.e. undergoing late-time inflation) Robertson-Walker spacetime and extra space with a constant 
scale factor.   In the case of multibrane solutions, some dimensions are static or shrink.  
  The fact that the largest supersymmetry algebra contains 32 supercharges in 4 dimensions 
imposes a restriction on the  geometry of extra space.
\\
keywords: Spacelike braneworld, Accelerating cosmologies
 }
\section{Introduction}

The first observational evidence that the present universe is still inflating (late-time inflation) 
was shown by Riess {\it et al.} \cite{Riess1} and Perlmutter {\it et al.} \cite{Perl}, separately.  
 Later observations \cite{cosmo, Riess2, Astier, Riess3} 
 supported this claim.  

If the cosmological constant is equal to zero or absorbed into $\rho$ and $P$, then the condition for inflation can be written as
\[
\rho + 3P < 0,
\]
where $\rho$ is the energy density of the universe and $P$ is its pressure.  Because the left-hand side of this equation is always positive 
for ordinary matter, radiation and dark matter, the existence of dark energy is regarded as indispensable in accounting for 
late-time inflation.  Although we have yet to unmask dark energy, 
some potential candidates, such as the cosmological constant and quintessence, have been proposed.  If we define the ratio of $P$ to $\rho$ as
\[
w \equiv {P\over \rho},
\]
then $w=-1$, if the cosmological constant is identified as dark energy.  On the other hand, $w$ usually fluctuates if quintessence assumptions are adopted.  
Recent observations \cite{Astier, Riess3, wmap, sdss} support the former candidate.  If the cosmological constant is the only source
  of late-time inflation, it will be a very small positive value, but not zero.  Although the cosmological constant can be chosen arbitrarily within 
  the limits of general relativity, it should be zero if general relativity is to be regarded as a low-energy effective theory of M-theory/superstring theory.


Many models of cosmological inflation have been posited based on M-theory/superstring theory. 
 However, one no-go theorem \cite{nogo} suggests that 4-dimensional de Sitter spacetime could not realized by the ordinary compactification  methods 
available under M theory/superstring theory.  
One way to overcome this problem \cite{ngreview} would be to put S-branes, which are time-dependent spacelike branes, into the model 
\cite{single, sbrane, sbrane2, sbrane3, sbrane4}.  
As D{\it p}-branes are objects which extend in {\it p} space dimensions and a time dimension, S{\it q}-branes extend in ($q+1$) space dimensions 
in ordinary notation.  It is known that accelerating S-brane solutions exist and that eternal inflation is possible if both internal and external spaces are 
hyperbolic \cite{ohta, eternal}.  Higher-order quantum corrections have been added to the action, 
and  solutions which include exponentially expanding braneworld and 
static extra
 space have been found numerically \cite{quan, quan2}.  In a recent paper \cite{mochi}, we pointed out the existence of 
 singular single-brane solutions to Einstein gravity coupled to a dilaton and an $n$-field.  One of these analytic solutions includes a 4-dimensional 
 exponentially expanding Robertson-Walker spacetime consisting of an SM2-brane and a time axis, and an 
 extra hyperbolic space $H^{D-p-2}$ with a constant scale factor.

In this paper, our starting point is Einstein gravity coupled to a dilaton and one or some $n$-form fields, 
 which yield multibrane solutions, as a low-energy effective theory of 
M-theory/superstring theory.    In order to obtain singular solutions, 
we set the time axis on the scale parameter of extra space, which may include more than one timelike metric, in 
contrast to our previous paper, in which only one timelike metric was considered.   
We obtain eternally inflating (i.e. undergoing late-time inflation) Robertson-Walker spacetime and extra space with a constant 
scale factor.   With multibrane solutions, some dimensions are static or shrink.  
The required intersection rule is shown to be the same as that suggested for regular solutions 
in other papers \cite{int1, int2, gen, multi}.  The last section is devoted to an examination of 
these solutions in relation to the real Universe.  The fact that the largest supersymmetry algebra contains 32 supercharges in 4 dimensions 
imposes a restriction on the geometry of extra space.


\section{Intersecting S-brane solution}

We consider Einstein gravity coupled to a dilaton field $\phi$ and  $m$ kinds of {\it n}-form fields $F_{n}$ as a low-energy effective theory of 
M-theory/superstring theory, whose action $I$ is

\begin{equation}
I={1\over 16\pi G}\int d^Dx\sqrt{\pm g}\Big[ R-{1\over 2}g^{\mu\nu}\partial_{\mu} \phi\partial_{\nu}\phi
-\sum_{A=1}^{m}{1\over 2\cdot n_A!}e^{\alpha_A\phi}F_{n_A}^{\ 2}\Big],\label{eq:action}
\end{equation}
where the sign  in front of $g$ should be selected as appropriate.  $\alpha_A$ is the dilaton coupling constant given by
\[
\alpha_A = \left\{
\begin{array}{cl}
0 & ({\rm M-theory}) \\
-1 & ({\rm NS-NS\  sector}) \\
{5-n_A\over 2} & ({\rm R-R\  sector}),
\end{array}
\right.
\]
and the bare cosmological constant is assumed to be zero.  {\it D}=11
 for M-theory and {\it D}=10 for superstring theories.

We assume the following metric form:
\begin{equation}
ds^2=\sum_{i=1}^{p+1}e^{2u_i}dx^idx^i+\sum_{a,b=p+2}^D e^{2v}\eta_{ab}dy^ady^b,\label{eq:setting}
\end{equation}
where
\begin{equation}
\eta_{ab} = \{diag.(\underbrace{+ , \cdots , +}_{D - p - 1 - s} , \underbrace{- , \cdots , -}_{s})\}.\label{eq:singletime}
\end{equation}
We use $x^i,\ i=1,\cdots,\ {\rm and}\ p+1$ as the coordinates of the space where the branes exist.   
A recent study has posited  singular single-brane solutions for the case where only one minus sign is to be included in the metric \cite{mochi}.  
 General orthogonally intersecting solutions where the metric functions $u$ and $v$ 
 and fields $\phi$ and $F$ depend only on one timelike coordinate have also been proposed \cite{gen}.   A D-brane solution which depends on all the extra space coordinates 
 has been suggested \cite{multi}.
 
  We assume that the metrics and fields do not depend only on the timelike coordinates $y^a$, $a=D-s+1, \cdots ,\ {\rm and}\  D$, 
 but also on the other perpendicular coordinates $y^a,\ a=p+2,\cdots,\ {\rm and}\ D-s$.  That is to say,
 \[
u = u(y) \equiv u(y^{p+2},\cdots,y^D),
\]
\[
v = v(y) \equiv v(y^{p+2},\cdots,y^D),
\]
\[
\phi = \phi(y) \equiv \phi(y^{p+2},\cdots,y^D),
\]
\[
F = F(y) \equiv F(y^{p+2},\cdots,y^D).
\]

The field strength for an electrically charged S{\it p}-brane is given by
\begin{equation}
(F_n)_{i_{1}\cdots i_{n-1}a}(y)=\epsilon_{i_1\cdots i_{n-1}}\partial_a E(y),
\end{equation}
where
\[
n = q+2.
\]
The magnetically charged case is given by
\begin{equation}
(F_n)^{a_1\cdots a_{n}}=\frac{1}{\sqrt{\pm g}}e^{-\alpha\phi}\epsilon^{a_1\cdots a_{n}b}\partial_b E(y),
\end{equation}
where
\[
n = D-q-2.
\]

The field equations are
\begin{eqnarray}
-\partial^2 u_i-(\partial u_i)\big\{\sum_{k=1}^{p+1}(\partial u_k)+(D-p-3)(\partial v)\big\}\nonumber\\
 = \sum_{A=1}^m{\delta_{A,i}\over 2(D-2)}e^{\varepsilon_A\alpha_A\phi-2\sum_{k\in qA}u_k}(\partial E_A)^2,\label{eq:eq1}
\end{eqnarray}
\begin{eqnarray}
\eta_{ab}\Big[-\partial^2 v-(\partial v)\big\{(D-p-3)(\partial v)+\sum_{k=1}^{p+1}(\partial u_k)\big\}\Big]  - \sum_{k=1}^{p+1}\partial_au_k\partial_bu_k\nonumber\\
 - \partial_a\partial_b\Big[(D-p-3)v + \sum_{k=1}^{p+1}u_k\Big] - (D-p-3)\partial_av\partial_bv\nonumber\\
 + \partial_av\Big[(D-p-3)\partial_bv + \sum_{k=1}^{p+1}\partial_bu_k\Big] + (a \leftrightarrow  b)\nonumber \\
=\sum_{A=1}^m{1\over 2} e^{\varepsilon_A\alpha_A\phi-2\sum_{k\in qA}u_k}\Big[\partial_aE_A\partial_bE_A-{q_A+1\over D-2} \eta_{ab}(\partial E_A)^2\Big]\nonumber\\
 + {1\over 2}\partial_a\phi\partial_b\phi,
\end{eqnarray}
\begin{eqnarray}
\partial_a\Big[e^{\sum_{k=1}^{p+1}u_k+(D-p-3)v} \eta^{ab} \partial_b\phi\Big]\nonumber\\
=\sum_{A=1}^m{\varepsilon_A\alpha_A \over 2}
e^{\varepsilon_A\alpha_A\phi-\sum_{k\in qA}u_k+\sum_{j\notin qA}u_j+(D-p-3)v} (\partial E_A)^2,
\end{eqnarray}
\begin{equation}
\partial_a\Big[e^{\varepsilon_A\alpha_A\phi-\sum_{k\in qA}u_k+\sum_{j\notin qA}u_j+(D-p-3)v} \eta^{ab} \partial_bE_A\Big]=0,\label{eq:eq2}
\end{equation}
and the Bianchi identity is 
\begin{equation}
\partial_{[a}F_{\cdots ]} = 0.\label{eq:bianchi}
\end{equation}
In these equations, we have introduced notations defined as
\[
\eta^{ab}\eta_{bc} = \delta^a_c
\]
\[
\partial^2 \equiv \eta^{ab}\partial_a\partial_b,
\]
\[
(\partial f)(\partial g) \equiv \eta^{ab}(\partial_a f)(\partial_b g)
\]
\[
\varepsilon_A = \left\{
\begin{array}{cl}
+1 & (F_{n_A}\ {\rm is\ an\ electric\ field\ strength}) \\
-1 & (F_{n_A}\ {\rm is\ a\ magnetic\ field\ strength})
\end{array}
\right.
\]
\[
\delta_{A,i} = \left\{
\begin{array}{cl}
D-q_A-3 & (i\in q_A) \\
-(q_A+1) & (i\notin q_A).
\end{array}
\right.
\]

To simplify the calculation, we assume 
\begin{equation}
\sum_{k=1}^{p+1}u_k(y)+(D-p-3)v(y)=0.\label{eq:assumption}
\end{equation}
Then, the above field equations and the Bianchi identity become
\begin{equation}
-\partial^2u_i = \sum_{A=1}^m{\delta_{A,i}\over 2(D-2)} e^{\varepsilon_A\alpha_A\phi-2\sum_{k\in qA}u_k} (\partial E_A)^2,\label{eq:equ}
\end{equation}
\begin{eqnarray}
-\eta_{ab}\partial^2 v  - \sum_{k=1}^{p+1}\partial_au_k\partial_bu_k - (D-p-3)\partial_av\partial_bv\nonumber \\
=\sum_{A=1}^m{1\over 2} e^{\varepsilon_A\alpha_A\phi-2\sum_{k\in qA}u_k}\Big[\partial_aE_A\partial_bE_A-{q_A+1\over D-2} \eta_{ab}(\partial E_A)^2\Big]\nonumber\\
 + {1\over 2}\partial_a\phi\partial_b\phi,
\end{eqnarray}
\begin{equation}
\partial^2\phi={\varepsilon_A\alpha_A\over 2} e^{\varepsilon_A\alpha_A\phi-2\sum_{k\in qA}u_k}(\partial E_A)^2,\label{eq:eqphi}
\end{equation}
\begin{equation}
\partial_a\Big[e^{\varepsilon_A\alpha_A\phi-2\sum_{k\in qA}u_k}\eta^{ab}\partial_bE_A\Big]=0.\label{eq:eqE}
\end{equation}
The Bianchi identity (\ref{eq:bianchi}) is trivially satisfied in the electric case, as is the field equation (\ref{eq:eq2}) in the magnetic case.

We write candidates  for the solution of the above equations (\ref{eq:equ}) $\sim$ 
(\ref{eq:eqE}) which satisfy assumption (\ref{eq:assumption}) as
\begin{equation}
E_A(y)=iH_A(y),\label{eq:solE}
\end{equation}
\begin{equation}
u_i(y)=\sum_{A=1}^m{\delta_{A,i}\over 2(D-2)}\ln H_A(y),\label{eq:solu}
\end{equation}
\begin{equation}
v(y)=\sum_{A=1}^m{-(q_A+1)\over 2(D-2)}\ln H_A(y),\label{eq:solv}
\end{equation}
\begin{equation}
\phi(y) =\sum_{A=1}^m{-\varepsilon_A\alpha_A\over 2}\ln H_A(y),\label{eq:solphi}
\end{equation}
where
\begin{equation}
H_A(y) \equiv {Q_A\over h(y)},
\end{equation}
with $Q_A$ as a constant.

Substituting (\ref{eq:solE}) $\sim$ (\ref{eq:h}) into (\ref{eq:equ}) $\sim$ (\ref{eq:eqE}), we find sufficient conditions
\begin{equation}
\partial^2h(y)=0\label{eq:h}
\end{equation}
and 
\begin{equation}
-\varepsilon_A\varepsilon_B\alpha_A\alpha_B-2(\bar q+1)+{2(q_A+1)(q_B+1)\over D-2}=0,\label{eq:intersection}
\end{equation}
where $\bar q+1$ represents dimensions on which $q_A$-brane and $q_B$-brane cross.  (\ref{eq:intersection}) is the intersection rule, which 
has already been suggested for regular solutions in other papers \cite{int1, int2, gen, multi}.


\section{Eternally inflating braneworld}
Now, consider a metric which depends only on the scale parameter $r$ of either the entire, or a part of, the spacetime perpendicular to the brane:
\begin{equation}
r \equiv \sqrt{-\eta_{ab}y^{a}y^{b}},\ \ \ \ -\eta_{ab}y^{a}y^{b}>0.
\end{equation}
Note that $r$ is a timelike coordinate.  To satisfy (\ref{eq:h}),
\begin{equation}
h= r^{-(D-p-3)}.\label{eq:HH}
\end{equation}
Then, the metric of this spacetime is 
\begin{equation}
ds^2=\sum_{i=1}^{p+1}e^{2u_i}dx^idx^i-e^{2v}\Big(dr^2 - r^2d\Sigma^2_{D-p-2} \Big), \label{eq:metric}
\end{equation}
where $d\Sigma_{D-p-2}$ is the line element of a hypersurface with a unit scale factor.   
If $s = 1$ in (\ref{eq:singletime}), this hypersurface is a $(D-p-2)$-dimensional hyperbolic space $H^{D-p-2}$ with a unit scale factor.
  Meanwhile, it becomes a $(D-p-2)$-dimensional unit {\it timelike} spherical surface $-S^{D-p-2}$ if $s=D-p-1$.

We define cosmic time $t$ as
\begin{eqnarray}
dt&\equiv&e^{v}dr\nonumber\\
&=&Cr^{(D-p-3)\sum{-(q_A+1)\over 2(D-2)}}dr,
\end{eqnarray}
where
\begin{equation}
C=\prod_{A=1}^m Q_A^{{-(q_A+1)\over 2(D-2)}}.
\end{equation} 
It is most interesting that the case 
\begin{equation}
(D-p-3)\sum_{A=1}^m{(q_A+1)\over 2(D-2)}=1\label{eq:joukenn}
\end{equation}
is satisfied.  In this case, as
\begin{equation}
t= C\ln r,
\end{equation}
and
\begin{equation}
e^{2v}=C^2r^{-2},
\end{equation}
the metric (\ref{eq:metric}) becomes
\begin{equation}
ds^2=-dt^2+\sum_{i=1}^{p+1}e^{2u_i}dx^idx^i+C^2d\Sigma^2_{D-p-2}\ . \label{eq:metric2}
\end{equation}
Note that the scale factor of the extra space is independent of $t$ in this metric.  
On the other hand, because
\begin{equation}
e^{2u_i} = \Big(\prod_{A=1}^m Q_A^{\delta_{A,i}\over D-2}\Big)e^{{D-p-3\over C(D-2)}\sum_A \delta_{A,i}t},\label{eq:ui}
\end{equation}
the $i$-th dimension exponentially expands if $\sum_A \delta_{A,i}$ is positive, is static if $\sum_A \delta_{A,i}=0$, or 
shrinks if $\sum_A \delta_{A,i}$ is negative. 

The solutions satisfying  the above condition (\ref{eq:joukenn}) are given in the tables in Subsections 3.1, 3.2 and 
3.3 for M-theory, IIA superstring theory and IIB superstring theory, respectively.  In these tables, 
we use the following symbols:
\begin{center}
\begin{tabular}{|c|c|} \hline
symbol& dimension \\ \hline
$\oplus$ &exponentially expands \\ 
$\ominus$ & exponentially shrinks \\ 
$\odot$ & static \\ 
$\bullet$ & constitutes cosmic time and static extra space \\ 
$\circ$ & exponentially shrinks  \\ \hline
\end{tabular}
\end{center}


\subsection{M-theory}
　
\begin{center}1 brane\\
\begin{tabular}{|c|c|c|c|c|c|c|c|c|c|c|c|} \hline
{\rm M2} &$\oplus$ &$\oplus$ &$\oplus$  &  &  & & & & & & \\
Other dim. & & & &$\bullet$ &$\bullet$ &$\bullet$ &$\bullet$ &$\bullet$ &$\bullet$ &$\bullet$ &$\bullet$ \\ \hline
\end{tabular}
\end{center}

\begin{center}
\begin{tabular}{|c|c|c|c|c|c|c|c|c|c|c|c|} \hline
{\rm M5} &$\oplus$ &$\oplus$ &$\oplus$  &$\oplus$  &$\oplus$  &$\oplus$ & & & & & \\
Other dim. & & & & & & &$\bullet$ &$\bullet$ &$\bullet$ &$\bullet$ &$\bullet$ \\ \hline
\end{tabular}
\end{center}

\begin{center}2 branes\\
\begin{tabular}{|c|c|c|c|c|c|c|c|c|c|c|c|} \hline
{\rm M2} & $\oplus$ & $\oplus$ & $\oplus$ & & & & & & & &  \\
{\rm M2} &$\oplus$ & &  & $\oplus$ & $\oplus$ & & & & & & \\
Other dim. & & & & & &$\circ$ & $\bullet$ &$\bullet$ &$\bullet$ &$\bullet$ &$\bullet$\\ \hline
\end{tabular}
\end{center}

\begin{center}
\begin{tabular}{|c|c|c|c|c|c|c|c|c|c|c|c|} \hline
{\rm M2} & $\oplus$ & $\oplus$ & $\odot$ & & & & & & & &  \\
{\rm M5} & $\oplus$ & $\oplus$ & & $\odot$  & $\odot$ & $\odot$ & $\odot$ & & & & \\
Other dim. & & & & & & & &$\bullet$ &$\bullet$ &$\bullet$ &$\bullet$\\ \hline
\end{tabular}
\end{center}

\begin{center}3 branes\\
\begin{tabular}{|c|c|c|c|c|c|c|c|c|c|c|c|} \hline
{\rm M2} & $\oplus$ & $\odot$ & $\odot$ & & & & & & & &  \\
{\rm M2} & $\oplus$ & & &$\odot$ &$\odot$ & & & & & &  \\
{\rm M2} & $\oplus$ & &  & & & $\odot$ &$\odot$  & & & & \\
Other dim. & & & & & & & &$\bullet$ &$\bullet$ &$\bullet$ &$\bullet$\\ \hline
\end{tabular}
\end{center}

\begin{center}
\begin{tabular}{|c|c|c|c|c|c|c|c|c|c|c|c|} \hline
{\rm M2} & $\oplus$ & $\oplus$ & & $\odot$ & & & & & &  & \\
{\rm M2} & $\oplus$ & & $\oplus$ & & $\odot$ & & & & &  & \\
{\rm M2} & & $\oplus$ & $\oplus$ & & & $\odot$ & & & &  & \\
Other dim. & & & & & & & $\circ$ & $\bullet$ & $\bullet$ & $\bullet$ & $\bullet$\\ \hline
\end{tabular}
\end{center}

\begin{center}
\begin{tabular}{|c|c|c|c|c|c|c|c|c|c|c|c|} \hline
{\rm M5} & $\oplus$ & $\oplus$ & $\odot$ & $\odot$ & $\odot$ & $\odot$ & & & & &  \\
{\rm M5} & $\oplus$ & $\oplus$ & $\odot$ & $\odot$ & & & $\odot$ & $\odot$ & & &  \\
{\rm M5} & $\oplus$ & $\oplus$ & &  & $\odot$ & $\odot$ & $\odot$  &$\odot$ & & & \\
Other dim. & & & & & & & & &$\bullet$ &$\bullet$ &$\bullet$\\ \hline
\end{tabular}
\end{center}

\begin{center}4 branes\\
\begin{tabular}{|c|c|c|c|c|c|c|c|c|c|c|c|} \hline
{\rm M2} & $\oplus$ & $\odot$ & $\odot$ & & & & & & & &  \\
{\rm M2} & $\oplus$ & & & $\odot$ & $\odot$ & & & & & &  \\
{\rm M5} & $\oplus$ & $\odot$ & & $\odot$ & & $\odot$ & $\odot$ & $\odot$ & & &  \\
{\rm M5} & $\oplus$ & & $\odot$ &  & $\odot$ & $\odot$ & $\odot$  &$\odot$ & & & \\
Other dim. & & & & & & & & &$\bullet$ &$\bullet$ &$\bullet$\\ \hline
\end{tabular}
\end{center}

\begin{center}
\begin{tabular}{|c|c|c|c|c|c|c|c|c|c|c|c|} \hline
{\rm M2} & $\oplus$ & $\oplus$ & & $\odot$ & & & & & & &  \\
{\rm M2} & $\oplus$ & & $\oplus$ & & $\odot$ & & & & & &  \\
{\rm M5} & & $\oplus$ & $\oplus$ & $\odot$ & $\odot$ & $\odot$ & $\odot$ & & & &  \\
{\rm M5} & $\oplus$ & $\oplus$ & $\oplus$ & & & $\odot$ & $\odot$ & $\ominus$ & & &  \\ 
Other dim. & & & & & & & & & $\bullet$ & $\bullet$ & $\bullet$\\ \hline
\end{tabular}
\end{center}

\begin{center}5 branes\\
\begin{tabular}{|c|c|c|c|c|c|c|c|c|c|c|c|} \hline
{\rm M2} & $\oplus$ & $\oplus$ & & & $\odot$ & & & & & &  \\
{\rm M2} & & & $\oplus$ & $\oplus$ & $\odot$ & & & & & &  \\
{\rm M2} & $\oplus$ & & $\oplus$ & & &$\odot$ & & & & &  \\
{\rm M2} & & $\oplus$ & & $\oplus$ & & $\odot$ & & & & &  \\
{\rm M5} & $\oplus$ & $\oplus$ & $\oplus$ & $\oplus$  & & &$\ominus$  &$\ominus$ & & & \\
Other dim. & & & & & & & & &$\bullet$ &$\bullet$ &$\bullet$\\ \hline
\end{tabular}
\end{center}

\begin{center}6 branes\\
\begin{tabular}{|c|c|c|c|c|c|c|c|c|c|c|c|} \hline
{\rm M2} & $\oplus$ & $\oplus$ & & & $\odot$ & & & & & &  \\
{\rm M2} & $\oplus$ & & $\oplus$ & & &$\odot$ & & & & &  \\
{\rm M2} & $\oplus$ & & & $\oplus$ & & & $\odot$ & & & &  \\
{\rm M2} & & & $\oplus$ & $\oplus$ & $\odot$ & & & & & &  \\
{\rm M2} & & $\oplus$ & & $\oplus$ & &$\odot$ & & & & &  \\
{\rm M2} & & $\oplus$ & $\oplus$ & & & & $\odot$ & & & &  \\
Other dim. & & & & & & & &$\circ$ &$\bullet$ &$\bullet$ &$\bullet$\\ \hline
\end{tabular}
\end{center}

\subsection{IIA superstring theory}

\begin{center}2 branes\\
\begin{tabular}{|c|c|c|c|c|c|c|c|c|c|c|} \hline
{\rm D0} & $\oplus$ & & & & & & & & &  \\
{\rm D2} & & $\oplus$ & $\oplus$  & $\oplus$ & & & & & & \\
Other dim. & & & & &$\bullet$ &$\bullet$ &$\bullet$ &$\bullet$ &$\bullet$ &$\bullet$\\ \hline
\end{tabular}
\end{center}

\begin{center}
\begin{tabular}{|c|c|c|c|c|c|c|c|c|c|c|} \hline
{\rm F1} & $\oplus$ & $\oplus$ & & & & & & & &  \\
{\rm NS5} &$\oplus$ & $\oplus$ & $\odot$  & $\odot$ & $\odot$ & $\odot$ & & & & \\
Other dim. & & & & & & &$\bullet$ &$\bullet$ &$\bullet$ &$\bullet$\\ \hline
\end{tabular}
\end{center}

\begin{center}
\begin{tabular}{|c|c|c|c|c|c|c|c|c|c|c|} \hline
{\rm D2} & $\oplus$ & $\oplus$ & $\odot$ & & & & & & &  \\
{\rm D4} &$\oplus$ & $\oplus$ &  & $\odot$ & $\odot$ & $\odot$ & & & & \\
Other dim. & & & & & & &$\bullet$ &$\bullet$ &$\bullet$ &$\bullet$\\ \hline
\end{tabular}
\end{center}

\begin{center}3 branes\\
\begin{tabular}{|c|c|c|c|c|c|c|c|c|c|c|} \hline
{\rm D0} & $\oplus$ & & & & & & & & &  \\
{\rm F1} & $\oplus$ & $\odot$ & & & & & & & &  \\
{\rm D4} & $\oplus$ & & $\odot$  & $\odot$ & $\odot$ & $\odot$ & & & & \\
Other dim. & & & & & & &$\bullet$ &$\bullet$ &$\bullet$ &$\bullet$\\ \hline
\end{tabular}
\end{center}

\begin{center}
\begin{tabular}{|c|c|c|c|c|c|c|c|c|c|c|} \hline
{\rm F1} & $\oplus$ & $\odot$ & & & & & & & &  \\
{\rm D2} & $\oplus$ & &$\odot$ &$\odot$ & & & & & &  \\
{\rm D2} & $\oplus$ & &  & & $\odot$ & $\odot$ & & & & \\
Other dim. & & & & & & &$\bullet$ &$\bullet$ &$\bullet$ &$\bullet$\\ \hline
\end{tabular}
\end{center}

\begin{center}
\begin{tabular}{|c|c|c|c|c|c|c|c|c|c|c|} \hline
{\rm F1} & $\oplus$ & $\oplus$ & & & & & & & & \\
{\rm D2} & $\oplus$ & & $\oplus$ & $\odot$ & & & & & & \\
{\rm D2} & & $\oplus$ & $\oplus$ & & $\odot$ & & & & & \\
Other dim. & & & & & & $\circ$ & $\bullet$ & $\bullet$ & $\bullet$ & $\bullet$\\ \hline
\end{tabular}
\end{center}

\begin{center}
\begin{tabular}{|c|c|c|c|c|c|c|c|c|c|c|} \hline
{\rm D4} & $\oplus$ & $\oplus$ & $\odot$ & $\odot$ & $\odot$ & & & & &  \\
{\rm D4} & $\oplus$ & $\oplus$ &$\odot$ & & &$\odot$ &$\odot$  & & &  \\
{\rm NS5} & $\oplus$ & $\oplus$ &  &$\odot$  & $\odot$ & $\odot$ &$\odot$  & & & \\
Other dim. & & & & & & & &$\bullet$ &$\bullet$ &$\bullet$\\ \hline
\end{tabular}
\end{center}

\begin{center}
\begin{tabular}{|c|c|c|c|c|c|c|c|c|c|c|} \hline
{\rm D2} & $\oplus$ & $\oplus$ & $\odot$ & & & & & & &  \\
{\rm NS5} & $\oplus$ & $\oplus$ &  &$\odot$  & $\odot$ & $\odot$ &$\odot$  & & & \\
{\rm D6} & $\oplus$ & $\oplus$ &$\odot$ & $\odot$ & $\odot$ &$\odot$ &$\odot$  & & &  \\
Other dim. & & & & & & & &$\bullet$ &$\bullet$ &$\bullet$\\ \hline
\end{tabular}
\end{center}

\begin{center}4 branes\\
\begin{tabular}{|c|c|c|c|c|c|c|c|c|c|c|} \hline
{\rm D0} & $\oplus$ & & & & & & & & &  \\
{\rm D4} & $\oplus$ & $\odot$ & $\odot$ & $\odot$ & $\odot$ & & & & &  \\
{\rm D4} & $\oplus$ & $\odot$ & $\odot$ & & & $\odot$ & $\odot$ & & &  \\
{\rm D4} & $\oplus$ & & & $\odot$ & $\odot$ & $\odot$ & $\odot$ & & &  \\
Other dim. & & & & & & & & $\bullet$ & $\bullet$ & $\bullet$\\ \hline
\end{tabular}
\end{center}

\begin{center}
\begin{tabular}{|c|c|c|c|c|c|c|c|c|c|c|} \hline
{\rm F1} & $\oplus$ & $\odot$ & & & & & & & &  \\
{\rm D2} & $\oplus$ & & $\odot$ & $\odot$ & & & & & &  \\
{\rm D4} & $\oplus$ & & $\odot$ & & $\odot$ & $\odot$ & $\odot$ & & &  \\
{\rm NS5} & $\oplus$ & $\odot$ &  & $\odot$ & $\odot$ & $\odot$  &$\odot$ & & & \\
Other dim. & & & & & & & &$\bullet$ &$\bullet$ &$\bullet$\\ \hline
\end{tabular}
\end{center}

\begin{center}
\begin{tabular}{|c|c|c|c|c|c|c|c|c|c|c|} \hline
{\rm F1} & $\oplus$ & $\oplus$ & & & & & & & &  \\
{\rm D2} & $\oplus$ & & $\oplus$ & $\odot$ & & & & & &  \\
{\rm D4} & & $\oplus$ & $\oplus$ & $\odot$ & $\odot$ & $\odot$ & & & &  \\
{\rm NS5} & $\oplus$ & $\oplus$ & $\oplus$ & & $\odot$ & $\odot$ & $\ominus$ & & &  \\
Other dim. & & & & & & & & $\bullet$ & $\bullet$ & $\bullet$\\ \hline
\end{tabular}
\end{center}

\begin{center}
\begin{tabular}{|c|c|c|c|c|c|c|c|c|c|c|} \hline
{\rm D2} & $\oplus$ & $\odot$ & $\odot$ & & & & & & &  \\
{\rm D2} & $\oplus$ & & & $\odot$ & $\odot$ & & & & &  \\
{\rm D2} & $\oplus$ & & & & & $\odot$ & $\odot$ & & &  \\
{\rm D6} & $\oplus$ & $\odot$ & $\odot$ & $\odot$  & $\odot$ & $\odot$ & $\odot$  & & & \\
Other dim. & & & & & & & &$\bullet$ &$\bullet$ &$\bullet$\\ \hline
\end{tabular}
\end{center}

\begin{center}
\begin{tabular}{|c|c|c|c|c|c|c|c|c|c|c|} \hline
{\rm D2} & $\oplus$ & $\oplus$ & & $\odot$ & & & & & &  \\
{\rm D2} & $\oplus$ & & $\oplus$ & & $\odot$ & & & & &  \\
{\rm D2} & & $\oplus$ & $\oplus$ & & & $\odot$ & & & &  \\
{\rm D6} & $\oplus$ & $\oplus$ & $\oplus$ & $\odot$ & $\odot$ & $\odot$ & $\ominus$ & & &  \\
Other dim. & & & & & & & & $\bullet$ & $\bullet$ & $\bullet$\\ \hline
\end{tabular}
\end{center}

\begin{center}
\begin{tabular}{|c|c|c|c|c|c|c|c|c|c|c|} \hline
{\rm D2} & $\oplus$ & $\odot$ & $\odot$ & & & & & & &  \\
{\rm D2} & $\oplus$ & & & $\odot$ & $\odot$ & & & & &  \\
{\rm D4} & $\oplus$ & $\odot$ & & $\odot$ & & $\odot$ & $\odot$ & & &  \\
{\rm D4} & $\oplus$ & & $\odot$ &  & $\odot$ & $\odot$ & $\odot$  & & & \\
Other dim. & & & & & & & &$\bullet$ &$\bullet$ &$\bullet$\\ \hline
\end{tabular}
\end{center}

\begin{center}
\begin{tabular}{|c|c|c|c|c|c|c|c|c|c|c|} \hline
{\rm D2} & $\oplus$ & $\oplus$ & & $\odot$ & & & & & &  \\
{\rm D2} & $\oplus$ & & $\oplus$ & & $\odot$ & & & & &  \\
{\rm D4} & & $\oplus$ & $\oplus$ & $\odot$ & $\odot$ & $\odot$ & & & &  \\
{\rm D4} & $\oplus$ & $\oplus$ & $\oplus$ & & & $\odot$ & $\ominus$ & & &  \\
Other dim. & & & & & & & & $\bullet$ & $\bullet$ & $\bullet$\\ \hline
\end{tabular}
\end{center}

\begin{center}5 branes\\
\begin{tabular}{|c|c|c|c|c|c|c|c|c|c|c|} \hline
{\rm F1} & $\oplus$ & $\oplus$ & & & & & & & &  \\
{\rm F1} & & & $\oplus$ & $\oplus$ & & & & & &  \\
{\rm D2} & $\oplus$ & & $\oplus$ & &$\odot$ & & & & &  \\
{\rm D2} & & $\oplus$ & & $\oplus$ & $\odot$ & & & & &  \\
{\rm NS5} & $\oplus$ & $\oplus$ & $\oplus$ & $\oplus$  & &$\ominus$  &$\ominus$ & & & \\
Other dim. & & & & & & & &$\bullet$ &$\bullet$ &$\bullet$\\ \hline
\end{tabular}
\end{center}

\begin{center}6 branes\\
\begin{tabular}{|c|c|c|c|c|c|c|c|c|c|c|} \hline
{\rm F1} & $\oplus$ & $\oplus$ & & & & & & & &  \\
{\rm F1} & & & $\oplus$ & $\oplus$ & & & & & &  \\
{\rm D2} & $\oplus$ & & $\oplus$ & & $\odot$ & &  & & &  \\
{\rm D2} & & $\oplus$ & & $\oplus$ & $\odot$ & & & & &  \\
{\rm D2} & & $\oplus$ & $\oplus$ & & & $\odot$ & & & &  \\
{\rm D2} & $\oplus$ & & & $\oplus$ & & $\odot$ & & & &  \\
Other dim. & & & & & & &$\circ$ &$\bullet$ &$\bullet$ &$\bullet$\\ \hline
\end{tabular}
\end{center}

\subsection{IIB superstring theory}

　
\begin{center}1 brane\\
\begin{tabular}{|c|c|c|c|c|c|c|c|c|c|c|} \hline
{\rm D3} &$\oplus$ &$\oplus$ &$\oplus$  & $\oplus$  &  & & & & & \\
Other dim. & & & & &$\bullet$ &$\bullet$ &$\bullet$ &$\bullet$ &$\bullet$ &$\bullet$ \\ \hline
\end{tabular}
\end{center}

\begin{center}2 branes\\
\begin{tabular}{|c|c|c|c|c|c|c|c|c|c|c|} \hline
{\rm F1} & $\oplus$ & $\oplus$ & & & & & & & &  \\
{\rm F1} & & & $\oplus$ & $\oplus$ & & & & & & \\
Other dim. & & & & & $\bullet$ & $\bullet$ &$\bullet$ &$\bullet$ &$\bullet$ &$\bullet$\\ \hline
\end{tabular}
\end{center}

\begin{center}
\begin{tabular}{|c|c|c|c|c|c|c|c|c|c|c|} \hline
{\rm D1} & $\oplus$ & $\oplus$ & & & & & & & &  \\
{\rm D1} & & & $\oplus$ & $\oplus$ & & & & & & \\
Other dim. & & & & & $\bullet$ & $\bullet$ &$\bullet$ &$\bullet$ &$\bullet$ &$\bullet$\\ \hline
\end{tabular}
\end{center}

\begin{center}
\begin{tabular}{|c|c|c|c|c|c|c|c|c|c|c|} \hline
{\rm F1} & $\oplus$ & $\oplus$ & & & & & & & &  \\
{\rm D1} & $\oplus$ & & $\oplus$ & & & & & & & \\
Other dim. & & & & $\circ$ & $\bullet$ & $\bullet$ &$\bullet$ &$\bullet$ &$\bullet$ &$\bullet$\\ \hline
\end{tabular}
\end{center}

\begin{center}
\begin{tabular}{|c|c|c|c|c|c|c|c|c|c|c|} \hline
{\rm F1} & $\oplus$ & $\oplus$ & & & & & & & &  \\
{\rm NS5} &$\oplus$ & $\oplus$ & $\odot$  & $\odot$ & $\odot$ & $\odot$ & & & & \\
Other dim. & & & & & & &$\bullet$ &$\bullet$ &$\bullet$ &$\bullet$\\ \hline
\end{tabular}
\end{center}

\begin{center}
\begin{tabular}{|c|c|c|c|c|c|c|c|c|c|c|} \hline
{\rm D1} & $\oplus$ & $\oplus$ & & & & & & & &  \\
{\rm D5} &$\oplus$ & $\oplus$ & $\odot$  & $\odot$ & $\odot$ & $\odot$ & & & & \\
Other dim. & & & & & & &$\bullet$ &$\bullet$ &$\bullet$ &$\bullet$\\ \hline
\end{tabular}
\end{center}

\begin{center}
\begin{tabular}{|c|c|c|c|c|c|c|c|c|c|c|} \hline
{\rm D3} & $\oplus$ & $\oplus$ & $\odot$ & $\odot$ & & & & & &  \\
{\rm D3} &$\oplus$ & $\oplus$ &  & & $\odot$ & $\odot$ & & & & \\
Other dim. & & & & & & &$\bullet$ &$\bullet$ &$\bullet$ &$\bullet$\\ \hline
\end{tabular}
\end{center}

\begin{center}3 branes\\
\begin{tabular}{|c|c|c|c|c|c|c|c|c|c|c|} \hline
{\rm F1} & $\oplus$ & & $\odot$ & & & & & & &  \\
{\rm F1} & & $\oplus$ & & $\odot$ & & & & & &  \\
{\rm D3} & $\oplus$ & $\oplus$ &  & & $\odot$ & $\odot$ & & & & \\
Other dim. & & & & & & &$\bullet$ &$\bullet$ &$\bullet$ &$\bullet$\\ \hline
\end{tabular}
\end{center}

\begin{center}
\begin{tabular}{|c|c|c|c|c|c|c|c|c|c|c|} \hline
{\rm F1} & $\oplus$ & $\odot$ & & & & & & & &  \\
{\rm D1} & $\oplus$ & & $\odot$ & & & & & & &  \\
{\rm D3} & $\oplus$ &  &  & $\odot$ & $\odot$ & $\odot$ & & & & \\
Other dim. & & & & & & &$\bullet$ &$\bullet$ &$\bullet$ &$\bullet$\\ \hline
\end{tabular}
\end{center}

\begin{center}
\begin{tabular}{|c|c|c|c|c|c|c|c|c|c|c|} \hline
{\rm F1} & $\oplus$ & $\oplus$ & & & & & & & & \\
{\rm D1} & $\oplus$ & & $\oplus$ & & & & & & & \\
{\rm D3} & & $\oplus$ & $\oplus$ & $\odot$ & $\odot$ & & & & & \\
Other dim. & & & & & & $\circ$ & $\bullet$ & $\bullet$ & $\bullet$ & $\bullet$\\ \hline
\end{tabular}
\end{center}

\begin{center}
\begin{tabular}{|c|c|c|c|c|c|c|c|c|c|c|} \hline
{\rm D1} & $\oplus$ & & $\odot$ & & & & & & &  \\
{\rm D1} & & $\oplus$ & & $\odot$ & & & & & &  \\
{\rm D3} & $\oplus$ & $\oplus$ &  & & $\odot$ & $\odot$ & & & & \\
Other dim. & & & & & & &$\bullet$ &$\bullet$ &$\bullet$ &$\bullet$\\ \hline
\end{tabular}
\end{center}

\begin{center}
\begin{tabular}{|c|c|c|c|c|c|c|c|c|c|c|} \hline
{\rm D3} & $\oplus$ & $\oplus$ & $\odot$ & $\odot$ & & & & & &  \\
{\rm NS5} & $\oplus$ & $\oplus$ &$\odot$ & & $\odot$ &$\odot$ &$\odot$  & & &  \\
{\rm D5} & $\oplus$ & $\oplus$ &  &$\odot$  & $\odot$ & $\odot$ &$\odot$  & & & \\
Other dim. & & & & & & & & $\bullet$ & $\bullet$ & $\bullet$\\ \hline
\end{tabular}
\end{center}

\begin{center}4 branes\\
\begin{tabular}{|c|c|c|c|c|c|c|c|c|c|c|} \hline
{\rm F1} & $\oplus$ & $\oplus$ & & & & & & & &  \\
{\rm F1} & & & $\oplus$ & $\oplus$ & & & & & &  \\
{\rm D1} & $\oplus$ & & $\oplus$ & & & & & & &  \\
{\rm D1} & & $\oplus$ & & $\oplus$ & & & & & &  \\
Other dim. & & & & & $\circ$ & $\circ$ & $\bullet$ & $\bullet$ & $\bullet$ & $\bullet$\\ \hline
\end{tabular}
\end{center}

\begin{center}
\begin{tabular}{|c|c|c|c|c|c|c|c|c|c|c|} \hline
{\rm F1} & $\oplus$ & $\odot$ & & & & & & & &  \\
{\rm D1} & $\oplus$ & & $\odot$ & & & & & & &  \\
{\rm NS5} & $\oplus$ & $\odot$ & & $\odot$ & $\odot$ & $\odot$ & $\odot$ & & &  \\
{\rm D5} & $\oplus$ & & $\odot$ & $\odot$ & $\odot$ & $\odot$ & $\odot$ & & &  \\
Other dim. & & & & & & & & $\bullet$ & $\bullet$ & $\bullet$\\ \hline
\end{tabular}
\end{center}

\begin{center}
\begin{tabular}{|c|c|c|c|c|c|c|c|c|c|c|} \hline
{\rm F1} & $\oplus$ & $\odot$ & & & & & & & &  \\
{\rm D3} & $\oplus$ & & $\odot$ & $\odot$ & $\odot$ & & & & &  \\
{\rm D3} & $\oplus$ & & $\odot$ & & & $\odot$ & $\odot$ & & &  \\
{\rm NS5} & $\oplus$ & $\odot$ & & $\odot$ & $\odot$ & $\odot$ & $\odot$ & & &  \\
Other dim. & & & & & & & & $\bullet$ & $\bullet$ & $\bullet$\\ \hline
\end{tabular}
\end{center}

\begin{center}
\begin{tabular}{|c|c|c|c|c|c|c|c|c|c|c|} \hline
{\rm F1} & $\oplus$ & $\oplus$ & & & & & & & &  \\
{\rm D3} & $\oplus$ & & $\oplus$ & $\odot$ & & $\odot$ & & & &  \\
{\rm D3} & & $\oplus$ & $\oplus$ & & $\odot$ & $\odot$ & & & &  \\
{\rm NS5} & $\oplus$ & $\oplus$ & $\oplus$ & $\odot$ & $\odot$ & & $\ominus$ & & &  \\
Other dim. & & & & & & & & $\bullet$ & $\bullet$ & $\bullet$\\ \hline
\end{tabular}
\end{center}

\begin{center}
\begin{tabular}{|c|c|c|c|c|c|c|c|c|c|c|} \hline
{\rm D1} & $\oplus$ & $\odot$ & & & & & & & &  \\
{\rm D3} & $\oplus$ & & $\odot$ & $\odot$ & $\odot$ & & & & &  \\
{\rm D3} & $\oplus$ & & $\odot$ & & & $\odot$ & $\odot$ & & &  \\
{\rm D5} & $\oplus$ & $\odot$ & & $\odot$ & $\odot$ & $\odot$ & $\odot$ & & &  \\
Other dim. & & & & & & & & $\bullet$ & $\bullet$ & $\bullet$\\ \hline
\end{tabular}
\end{center}

\begin{center}
\begin{tabular}{|c|c|c|c|c|c|c|c|c|c|c|} \hline
{\rm D1} & $\oplus$ & $\oplus$ & & & & & & & &  \\
{\rm D3} & $\oplus$ & & $\oplus$ & $\odot$ & & $\odot$ & & & &  \\
{\rm D3} & & $\oplus$ & $\oplus$ & & $\odot$ & $\odot$ & & & &  \\
{\rm D5} & $\oplus$ & $\oplus$ & $\oplus$ & $\odot$ & $\odot$ & & $\ominus$ & & &  \\
Other dim. & & & & & & & & $\bullet$ & $\bullet$ & $\bullet$\\ \hline
\end{tabular}
\end{center}

\begin{center}
\begin{tabular}{|c|c|c|c|c|c|c|c|c|c|c|} \hline
{\rm D3} & $\oplus$ & $\odot$ & $\odot$ & $\odot$ & & & & & &  \\
{\rm D3} & $\oplus$ & $\odot$ & & & $\odot$ & $\odot$ & & & &  \\
{\rm D3} & $\oplus$ & & $\odot$ & & $\odot$ & & $\odot$ & & &  \\
{\rm D3} & $\oplus$ & & & $\odot$ & & $\odot$ & $\odot$ & & &  \\
Other dim. & & & & & & & & $\bullet$ & $\bullet$ & $\bullet$\\ \hline
\end{tabular}
\end{center}

\begin{center}
\begin{tabular}{|c|c|c|c|c|c|c|c|c|c|c|} \hline
{\rm D3} & $\oplus$ & $\oplus$ & $\oplus$ & & & & $\ominus$ & & &  \\
{\rm D3} & $\oplus$ & $\oplus$ & & $\odot$ & $\odot$ & & & & &  \\
{\rm D3} & $\oplus$ & & $\oplus$ & $\odot$ & & $\odot$ & & & &  \\
{\rm D3} & & $\oplus$ & $\oplus$ & & $\odot$ & $\odot$ & & & &  \\
Other dim. & & & & & & & & $\bullet$ & $\bullet$ & $\bullet$\\ \hline
\end{tabular}
\end{center}

\begin{center}5 branes\\
\begin{tabular}{|c|c|c|c|c|c|c|c|c|c|c|} \hline
{\rm F1} & $\oplus$ & & & $\odot$ & & & & & &  \\
{\rm F1} & & $\oplus$ & & & $\odot$ & & & & &  \\
{\rm D3} & $\oplus$ & $\oplus$ & $\oplus$ & & & & $\ominus$ & & &  \\
{\rm D3} & $\oplus$ & & $\oplus$ & & $\odot$  & $\odot$ & & & &  \\
{\rm D3} & & $\oplus$ & $\oplus$ & $\odot$ & & $\odot$ & & & &  \\
Other dim. & & & & & & & & $\bullet$ & $\bullet$ & $\bullet$\\ \hline
\end{tabular}
\end{center}

\begin{center}
\begin{tabular}{|c|c|c|c|c|c|c|c|c|c|c|} \hline
{\rm D1} & $\oplus$ & & & $\odot$ & & & & & &  \\
{\rm D1} & & $\oplus$ & & & $\odot$ & & & & &  \\
{\rm D3} & $\oplus$ & $\oplus$ & $\oplus$ & & & & $\ominus$ & & &  \\
{\rm D3} & $\oplus$ & & $\oplus$ & & $\odot$  & $\odot$ & & & &  \\
{\rm D3} & & $\oplus$ & $\oplus$ & $\odot$ & & $\odot$ & & & &  \\
Other dim. & & & & & & & & $\bullet$ & $\bullet$ & $\bullet$\\ \hline
\end{tabular}
\end{center}

\begin{center}
\begin{tabular}{|c|c|c|c|c|c|c|c|c|c|c|} \hline
{\rm F1} & $\oplus$ & & & $\odot$ & & & & & &  \\
{\rm F1} & & $\oplus$ & & & $\odot$ & & & & &  \\
{\rm F1} & & & $\oplus$ & & & $\odot$ & & & &  \\
{\rm D3} & $\oplus$ & $\oplus$ & $\oplus$ & &  &  & $\ominus$ & & &  \\
{\rm NS5} & $\oplus$ & $\oplus$ & $\oplus$ & $\odot$ & $\odot$ & $\odot$ & & & &  \\
Other dim. & & & & & & & & $\bullet$ & $\bullet$ & $\bullet$\\ \hline
\end{tabular}
\end{center}

\begin{center}
\begin{tabular}{|c|c|c|c|c|c|c|c|c|c|c|} \hline
{\rm D1} & $\oplus$ & & & $\odot$ & & & & & &  \\
{\rm D1} & & $\oplus$ & & & $\odot$ & & & & &  \\
{\rm D1} & & & $\oplus$ & & & $\odot$ & & & &  \\
{\rm D3} & $\oplus$ & $\oplus$ & $\oplus$ & &  &  & $\ominus$ & & &  \\
{\rm D5} & $\oplus$ & $\oplus$ & $\oplus$ & $\odot$ & $\odot$ & $\odot$ & & & &  \\
Other dim. & & & & & & & & $\bullet$ & $\bullet$ & $\bullet$\\ \hline
\end{tabular}
\end{center}

\begin{center}6 branes\\
\begin{tabular}{|c|c|c|c|c|c|c|c|c|c|c|} \hline
{\rm F1} & $\oplus$ & $\oplus$ & & & & & & & &  \\
{\rm F1} & & & $\oplus$ & $\oplus$ & & & & & &  \\
{\rm D1} & $\oplus$ & & $\oplus$ & & & &  & & &  \\
{\rm D1} & & $\oplus$ & & $\oplus$ & & & & & &  \\
{\rm D3} & $\oplus$ & & & $\oplus$ & $\odot$ & $\odot$ & & & &  \\
{\rm D3} & & $\oplus$ & $\oplus$ & & $\odot$ & $\odot$ & & & &  \\
Other dim. & & & & & & &$\circ$ &$\bullet$ &$\bullet$ &$\bullet$\\ \hline
\end{tabular}
\end{center}

\section{Concluding remarks}
In the previous sections, we constructed intersecting spacelike braneworld models.  To
 examine the reality of these models, we focus on the following 9 models, which include  3 isotropically expanding space dimensions.  

From M-theory, 2 models are nominated:
\begin{center}
\begin{tabular}{|c|c|c|c|c|c|c|c|c|c|c|c|} \hline
{\rm M2} &$\oplus$ &$\oplus$ &$\oplus$  &  &  & & & & & & \\
Other dim. & & & &$\bullet$ &$\bullet$ &$\bullet$ &$\bullet$ &$\bullet$ &$\bullet$ &$\bullet$ &$\bullet$ \\ \hline
\end{tabular}
\end{center}

\begin{center}
\begin{tabular}{|c|c|c|c|c|c|c|c|c|c|c|c|} \hline
{\rm M2} & $\oplus$ & $\oplus$ & & $\odot$ & & & & & &  & \\
{\rm M2} & $\oplus$ & & $\oplus$ & & $\odot$ & & & & &  & \\
{\rm M2} & & $\oplus$ & $\oplus$ & & & $\odot$ & & & &  & \\
Other dim. & & & & & & & $\circ$ & $\bullet$ & $\bullet$ & $\bullet$ & $\bullet$\\ \hline
\end{tabular}
\end{center}
On the other hand, 3 and 7 candidates come from IIA and IIB superstring theories, respectively:

\begin{center}
\begin{tabular}{|c|c|c|c|c|c|c|c|c|c|c|} \hline
{\rm F1} & $\oplus$ & $\oplus$ & & & & & & & & \\
{\rm D2} & $\oplus$ & & $\oplus$ & $\odot$ & & & & & & \\
{\rm D2} & & $\oplus$ & $\oplus$ & & $\odot$ & & & & & \\
Other dim. & & & & & & $\circ$ & $\bullet$ & $\bullet$ & $\bullet$ & $\bullet$\\ \hline
\end{tabular}
\end{center}

\begin{center}
\begin{tabular}{|c|c|c|c|c|c|c|c|c|c|c|} \hline
{\rm D2} & $\oplus$ & $\oplus$ & & $\odot$ & & & & & &  \\
{\rm D2} & $\oplus$ & & $\oplus$ & & $\odot$ & & & & &  \\
{\rm D2} & & $\oplus$ & $\oplus$ & & & $\odot$ & & & &  \\
{\rm D6} & $\oplus$ & $\oplus$ & $\oplus$ & $\odot$ & $\odot$ & $\odot$ & $\ominus$ & & &  \\
Other dim. & & & & & & & & $\bullet$ & $\bullet$ & $\bullet$\\ \hline
\end{tabular}
\end{center}

\begin{center}
\begin{tabular}{|c|c|c|c|c|c|c|c|c|c|c|} \hline
{\rm D2} & $\oplus$ & $\oplus$ & & $\odot$ & & & & & &  \\
{\rm D2} & $\oplus$ & & $\oplus$ & & $\odot$ & & & & &  \\
{\rm D4} & & $\oplus$ & $\oplus$ & $\odot$ & $\odot$ & $\odot$ & & & &  \\
{\rm D4} & $\oplus$ & $\oplus$ & $\oplus$ & & & $\odot$ & $\ominus$ & & &  \\
Other dim. & & & & & & & & $\bullet$ & $\bullet$ & $\bullet$\\ \hline
\end{tabular}
\end{center}

\begin{center}
\begin{tabular}{|c|c|c|c|c|c|c|c|c|c|c|} \hline
{\rm F1} & $\oplus$ & $\oplus$ & & & & & & & & \\
{\rm D1} & $\oplus$ & & $\oplus$ & & & & & & & \\
{\rm D3} & & $\oplus$ & $\oplus$ & $\odot$ & $\odot$ & & & & & \\
Other dim. & & & & & & $\circ$ & $\bullet$ & $\bullet$ & $\bullet$ & $\bullet$\\ \hline
\end{tabular}
\end{center}

\begin{center}
\begin{tabular}{|c|c|c|c|c|c|c|c|c|c|c|} \hline
{\rm D1} & $\oplus$ & $\oplus$ & & & & & & & &  \\
{\rm D3} & $\oplus$ & & $\oplus$ & $\odot$ & & $\odot$ & & & &  \\
{\rm D3} & & $\oplus$ & $\oplus$ & & $\odot$ & $\odot$ & & & &  \\
{\rm D5} & $\oplus$ & $\oplus$ & $\oplus$ & $\odot$ & $\odot$ & & $\ominus$ & & &  \\
Other dim. & & & & & & & & $\bullet$ & $\bullet$ & $\bullet$\\ \hline
\end{tabular}
\end{center}

\begin{center}
\begin{tabular}{|c|c|c|c|c|c|c|c|c|c|c|} \hline
{\rm D3} & $\oplus$ & $\oplus$ & $\oplus$ & & & & $\ominus$ & & &  \\
{\rm D3} & $\oplus$ & $\oplus$ & & $\odot$ & $\odot$ & & & & &  \\
{\rm D3} & $\oplus$ & & $\oplus$ & $\odot$ & & $\odot$ & & & &  \\
{\rm D3} & & $\oplus$ & $\oplus$ & & $\odot$ & $\odot$ & & & &  \\
Other dim. & & & & & & & & $\bullet$ & $\bullet$ & $\bullet$\\ \hline
\end{tabular}
\end{center}

\begin{center}
\begin{tabular}{|c|c|c|c|c|c|c|c|c|c|c|} \hline
{\rm F1} & $\oplus$ & & & $\odot$ & & & & & &  \\
{\rm F1} & & $\oplus$ & & & $\odot$ & & & & &  \\
{\rm D3} & $\oplus$ & $\oplus$ & $\oplus$ & & & & $\ominus$ & & &  \\
{\rm D3} & $\oplus$ & & $\oplus$ & & $\odot$  & $\odot$ & & & &  \\
{\rm D3} & & $\oplus$ & $\oplus$ & $\odot$ & & $\odot$ & & & &  \\
Other dim. & & & & & & & & $\bullet$ & $\bullet$ & $\bullet$\\ \hline
\end{tabular}
\end{center}

\begin{center}
\begin{tabular}{|c|c|c|c|c|c|c|c|c|c|c|} \hline
{\rm D1} & $\oplus$ & & & $\odot$ & & & & & &  \\
{\rm D1} & & $\oplus$ & & & $\odot$ & & & & &  \\
{\rm D3} & $\oplus$ & $\oplus$ & $\oplus$ & & & & $\ominus$ & & &  \\
{\rm D3} & $\oplus$ & & $\oplus$ & & $\odot$  & $\odot$ & & & &  \\
{\rm D3} & & $\oplus$ & $\oplus$ & $\odot$ & & $\odot$ & & & &  \\
Other dim. & & & & & & & & $\bullet$ & $\bullet$ & $\bullet$\\ \hline
\end{tabular}
\end{center}

\begin{center}
\begin{tabular}{|c|c|c|c|c|c|c|c|c|c|c|} \hline
{\rm F1} & $\oplus$ & & & $\odot$ & & & & & &  \\
{\rm F1} & & $\oplus$ & & & $\odot$ & & & & &  \\
{\rm F1} & & & $\oplus$ & & & $\odot$ & & & &  \\
{\rm D3} & $\oplus$ & $\oplus$ & $\oplus$ & &  &  & $\ominus$ & & &  \\
{\rm NS5} & $\oplus$ & $\oplus$ & $\oplus$ & $\odot$ & $\odot$ & $\odot$ & & & &  \\
Other dim. & & & & & & & & $\bullet$ & $\bullet$ & $\bullet$\\ \hline
\end{tabular}
\end{center}

\begin{center}
\begin{tabular}{|c|c|c|c|c|c|c|c|c|c|c|} \hline
{\rm D1} & $\oplus$ & & & $\odot$ & & & & & &  \\
{\rm D1} & & $\oplus$ & & & $\odot$ & & & & &  \\
{\rm D1} & & & $\oplus$ & & & $\odot$ & & & &  \\
{\rm D3} & $\oplus$ & $\oplus$ & $\oplus$ & &  &  & $\ominus$ & & &  \\
{\rm D5} & $\oplus$ & $\oplus$ & $\oplus$ & $\odot$ & $\odot$ & $\odot$ & & & &  \\
Other dim. & & & & & & & & $\bullet$ & $\bullet$ & $\bullet$\\ \hline
\end{tabular}
\end{center}

Because each multi-brane model includes a shrinking space dimension, its classical dimension would reduce 
with the passage of time.  Therefore, change of the behavior of other dimensions might be caused.

First, we investigate the restrictions caused by the fact that the largest supersymmetry algebra in 4 dimensions is $N=8$ and that it contains 
32 (real) supercharges.  This is expected to hold  in our models.   In 11 dimensions, the number of real components of the smallest representation 
is 32, if the Majorana condition is satisfied.  Hence, the number of the timelike coordinates
 $s=1, 2, 5$ and $6$ is allowed in the above-mentioned M-theory model.  On the other hand, 
$s=1, 2$ and $4$ are allowed, as the Majorana 
condition or the Weyl condition must be satisfied in the superstring theory model.  If $s=1$, the extra space is a $(D-p-2)$-dimensional hyperbolic space 
$H^{D-p-2}$, 
whereas if $s=D-p-1$, it is a compact $(D-p-2)$-dimensional {\it timelike} sphere.   

Some studies 
  have pointed out the importance of compact hyperbolic manifolds for internal space.   Modding out $H^{D-p-2}$ by an appropriate 
  freely acting discrete subgroup of the 
isometry group of $H^{D-p-2}$, a compact hyperbolic manifold is obtained.  These studies posit a model in which the universe is the direct product of a Robertson-Walker 
spacetime and a compact hyperbolic manifold \cite{chm, chm2, chm3}.  

Secondly, we compare expansion of our braneworld with late-time inflation of the Universe.  For simplicity, we neglect a 
curvature term and think of an 
{\it effective} cosmological constant $\Lambda$ as the only content of the Universe. Then, the Friedmann equation of a $\Lambda$-dominated universe is 
\[
H^2 = {\Lambda \over 3},
\]
where $H$ is the present Hubble parameter 
\[
H \sim 10^{-26}{\rm m}^{-1}.
\]

Comparing its metric 
\[
ds^2 = - dt^2 + {\rm e}^{\sqrt{{\Lambda \over 3}}t}d{\bf x}^2
\]
with (\ref{eq:ui}), we obtain
\begin{equation}
\sqrt{{\Lambda\over 3}} = {D-p-3\over 2C(D-2)}\sum_A\delta_{A,i}\ .
\end{equation}
Because 
\[
{D-p-3\over 2(D-2)}\sum_A\delta_{A,i} \sim O(1),
\]
the scale factor of extra dimensions $C$ is estimated as
\begin{equation}
C \sim \Lambda^{-1/2}.
\end{equation}

Therefore,
\begin{equation}
C \sim H^{-1} \sim (\Lambda_{\rm late})^{-1/2} \sim 10^{26}{\rm m}.
\end{equation}
This value is comparable to the present size of the Universe.


\[
\]

We would like to thank Associate Professor Jeremy Williams, Tokyo Dental College, for his assistance with the English of this manuscript.

\end{document}